\documentstyle[11pt,newpasp,twoside,epsf]{article}
\def\edcomment#1{\iffalse\marginpar{\raggedright\sl#1\/}\else\relax\fi}
\reversemarginpar

\begin{document}
\title{PLANET~II:\\
A Microlensing and Transit Search for Extrasolar Planets}

\author{Penny D.~Sackett}
\affil{Research School of Astronomy and Astrophysics, ANU,
Mt. Stromlo Observatory, Cotter Road,
Weston Creek ACT 2611, Australia}

\author{M.D.~Albrow$^{1}$, J.-P.~Beaulieu$^{2}$,
J.A.R.~Caldwell$^{3}$, C.~Coutures$^{4}$, M.~Dominik,
J.~Greenhill$^{5}$, K.~Hill$^{5}$, K.~Horne$^{6}$,
U.-G.~Jorgensen$^{7}$, S.~Kane$^{6}$, D.~Kubas$^{8}$,
R.~Martin$^{9}$, J.W.~Menzies$^{10}$,
K.R.~Pollard$^{1}$,
K.C.~Sahu$^{3}$, J.~Wambsganss$^{8}$, R.~Watson$^{5}$,
A.~Williams$^{9}$ \\(The PLANET Collaboration)}

\affil{
$^1$University of Canterbury, Christchurch, New Zealand\\
$^2$Institut d'Astrophysique de Paris, Paris, France\\
$^3$Space Telescope Science Institute, Baltimore, Maryland, USA\\
$^4$European Southern Observatory, LaSilla, Chile\\
$^5$University of Tasmania, Hobart, Australia\\
$^6$St.~Andrews University, St.~Andrews, Scotland\\
$^7$Niels Bohr Institute, University of Copenhagen, Copenhagen, Denmark\\
$^8$University of Potsdam, Potsdam, Germany\\
$^{9}$Perth Observatory, Bickley, Australia\\
$^{10}$South African Astronomical Observatory, Cape Town, South Africa\\
}

\begin{abstract}
Due to their extremely small luminosity compared to the stars they orbit,
planets outside our own Solar System are extraordinarily difficult to
detect directly in optical light.  Careful photometric monitoring of distant
stars, however, can reveal the presence of exoplanets via the microlensing
or eclipsing effects they induce.  The international PLANET collaboration
is performing such monitoring using a cadre of semi-dedicated telescopes
around the world.  Their results constrain the number
of gas giants orbiting 1--7 AU from the most typical stars in the Galaxy.
Upgrades in the program are opening regions of ``exoplanet discovery space''
-- toward smaller masses and larger orbital radii -- that are inaccessible
to the Doppler velocity technique.
\end{abstract}

\section{Looking for Exoplanets through the Lens of Gravity}

The Doppler velocity technique, which measures the
small to and fro motion induced in a parent star by an orbiting
planet, has clearly demonstrated that several percent of solar-type stars
have planets very {\it unlike} those in the Solar System
(Marcy, Cochran \& Mayor 2000).
Known extrasolar planets are plotted with
the nine Solar System planets in Figure~1.
With the exception
of the few bodies orbiting dead stellar cores known as pulsars,
exoplanets discovered to date populate a portion of mass-orbit parameter
space that barely overlaps that of Solar System planets.  This
is due to selection effects of the Doppler method which is sensitive to
high-mass planets orbiting close to their parent bodies.  The best
Doppler searches have velocity precisions of 3 to 10~m/s, just below that needed
to detect a Jovian analog in the most favorable orientation.
A suite of search techniques is thus required
to explore the range of masses and orbital characteristics
exhibited by Solar System planets.
We describe a combined microlensing and transit search (PLANET II)
for extrasolar gas giants that begins this wider exploration.
A general
review of microlensing and transit
exoplanet searches is given by Sackett (1999).

\begin{figure}
\hglue 2.0cm\epsfxsize=9.1cm\epsffile{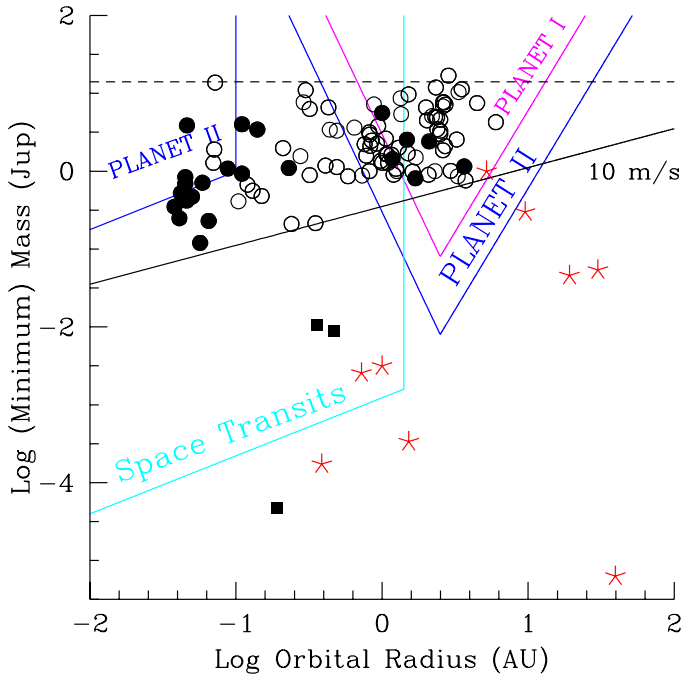}
\vglue -0.2cm
\caption{Solar system planets (stars) are shown on a log-log plot of
mass (Jupiter masses) vs semi-major axis (AU), with
known exoplanets orbiting pulsars
(squares) and main sequence stars (dots).
Open circles indicate exoplanets with orbits considerably more elongated
than those in the Solar System.
Planets creating a 10~m/s Doppler signal would lie on the indicated
solid line.
Discovery spaces for PLANET~I,
PLANET~II, and future space transit missions are sketched.
}
\vglue -0.2cm
\end{figure}

Gravitational microlensing occurs when a massive compact object (such as a star)
passes very near the line-of-sight to a background luminous source
(such as another star).  The gravitational field of the foreground
``lens'' bends the light rays from the background ``light bulb'', resulting
in more light reaching the observer's telescope.
Significant lensing requires precise source-lens alignment,
comparable to the angular radius of the so-called Einstein ring,
defined as
$\theta_{\rm E} \equiv [4 G M (1 - x)/(c^2 D_L)]^{1/2}$,
where $x \equiv D_L/D_S$, and $M$, $D_L$, and $D_S$ are the lens mass,
lens distance, and source star distance, respectively.
For typical Galactic microlensing, $ \theta_{\rm E} \approx$1~mas,
corresponding to separations of $1 - 5$~AU at the position of the
lenses.  Thus, when these rare precise alignments occur,
planetary orbits comparable in size to those
of Earth and Jupiter are detectable.

As an isolated lens, source, and observer move relative to one another,
different portions of the lens
magnification pattern are traversed by the source, resulting in a
symmetric rising and falling light curve.
If the lens is orbited
by another star or a planet, however, more complicated magnification patterns
are formed; the shape of the
resulting light curve then depends on the precise source trajectory
(Fig.~2), as described by Mao \& Paczyn\'ski (1991).
If the source passes near a defect -- or ``caustic'' --
in the lensing pattern caused by the companion, an anomalous
light curve will result that deviates from the smooth, symmetric
light curves of isolated lenses.  Precise characterization
of the anomaly allows the
mass ratio $q \equiv m_p/M$ and
projected separation $b$ of the lens and companion
(in units of $\theta_{\rm E}$) to be determined.
The size of caustics (and thus the probability
that a random background star will pass near one)
decreases slowly with mass ratio $q$ (Dominik 1999), making
microlensing an attractive method to search
for low-mass planets.

\begin{figure}
\vglue -7.7cm
\hglue 0.25cm\epsfxsize=12.5cm\epsffile{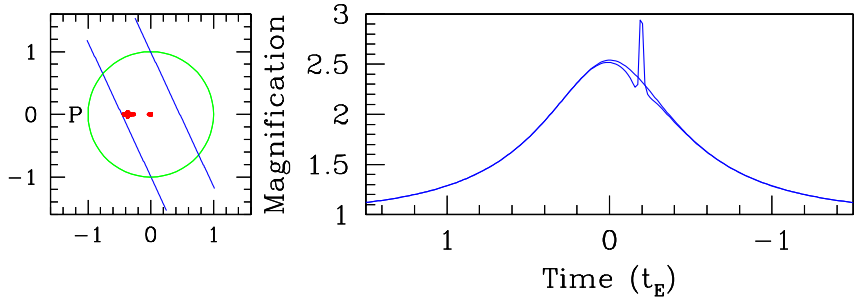}
\vglue -1.0cm
\caption{{\it Left:\/} The position of a Jovian
$q=10^{-3}$ planet is marked with ``P''
just outside the Einstein ring radius of its primary at $b=1.2$.  The
gravitational field of the primary lens is disturbed, creating
small ``caustics.''
{\it Right:\/} The resulting light curves from two possible source
trajectories shown at left. Only the light curve associated
with the path passing near a caustic
betrays the presence of the planet.
}
\vglue -0.2cm
\end{figure}

The precise alignment required for microlensing and the vast emptiness of
space conspire to reduce the probability of microlensing for background
Galactic stars at any given time
to $\sim$$10^{-6}$.  International groups such as
OGLE (Udalski et al.~1994) and MOA team (Bond et al.~2001)
monitor millions of
stars nightly in search of these needles in a haystack, issuing
real-time electronic alerts when they find them.
The Probing Lensing Anomalies NETwork (PLANET) then performs
the intensive and sensitive
photometric measurements necessary to detect low-mass lensing
companions.  Because planetary light curve anomalies
are short-lived (several hours to few days) and rare,
PLANET uses a network of semi-dedicated telescopes
longitudinally distributed around the Southern Hemisphere
for round-the-clock monitoring (Albrow et al.~1998).

Over its first five years, PLANET monitored
43 microlensing events intensely, but found no light curve anomalies
attributed to low-mass companions.  (Stellar binary systems, with
$q > 0.2$ were sometimes observed.)  By computing
the detection efficiency of each light curve
to planets with given $b$ and $q$ (Gaudi \& Sackett 2000), these null
results were translated into upper limits on
the frequency of gas giants orbiting the most typical stars
(ie., the lenses) in the Galaxy (Albrow et al.~2001; Gaudi et al.~2002).
No more than 50\% of the lenses can have planets with parameters
falling anywhere above the indicated PLANET~I line sketched in Fig.~1.
Specifically, less than 50\% of $\sim$0.3~M$_\odot$ stars have
companions with semi-major axes in the range
1.5~AU$\, < a < \, $4~AU and masses greater than that of Saturn.

\section{Expanding Discovery Space: PLANET~II}

The increased number of alerts now issued by OGLE and MOA
allow PLANET to select and monitor events that have particularly high
sensitivity to the presence of planets.  The overall planet detection
efficiency is expected to increase by a factor of
five, resulting in substantial sensitivity to Saturnian-mass planets
throughout the range $1$~AU$ < a < 10$~AU in orbital radius (Fig.~1).
The expanded program, PLANET~II, includes access to the
ESO~2.2m telescope; its wide field of view ($> 0.5$ degree on a side),
enables a simultaneous search
for transiting gas giants orbiting Solar-type
F and G stars in the same fields.
The size of the planet is determined from the depth of the partial
eclipse, the orbital period is given
by the time between successive transits.

\begin{figure}
\vglue -5.8cm
\hglue 0.85cm\epsfxsize=10.5cm\epsffile{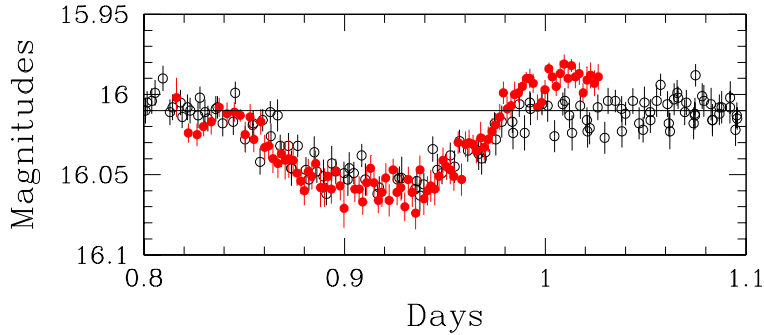}
\vglue -0.8cm
\caption{OGLE (open circles, Udalski et al. 2001) and
PLANET (filled dots, unpublished) data
for the transiting system OGLE-TR-18
(0.01mag$\approx$1\%).  The companion has a period of
2.228~days and is believed to have a radius twice that of Jupiter.
OGLE data were taken over six separate transits and then phase-wrapped;
PLANET data were taken with the ESO/Danish~1.5m telescope during
a single night.
}
\vglue -0.1cm
\end{figure}

The smaller the orbit, the larger the geometric probability of transiting.
So-called ``hot Jupiters,'' with periods of 3--5 days, have
$\sim$10\% chance of transiting solar-type parent stars, producing
photometric dips of amplitude $\sim$1\% lasting 1--3 hours.
About 1\% of solar neighborhood stars searched for Doppler signals
have such planets.
If hot Jupiters are
equally prevalent around inner Galaxy stars, simulations
suggest that up to 100 planets could be found per year by
searching for partial eclipses in wide-field surveys of the
Galaxy.  OGLE~III has already found tens of transit candidates
(Udalski et al.~2001);
recent test results that the image subtraction technique employed by
PLANET can generate comparable photometry (Fig.~3).
PLANET~II will sample fewer fields much more frequently and
with higher signal-to-noise when using the ESO~2.2m telescope.

Transit and microlensing searches probe two important
regimes of the angular momentum exchanging migration process
(Trilling et al.\ 1998, and
references therein) thought to be responsible for hot Jupiters.
Microlensing can
detect cool gas giants at the separations (several AU from the parent star)
where they are believed to have formed; transit photometry is
sensitive to hot gas giants that have migrated
during their infancy from these birthplaces to small, inner orbits.
Proper statistical analysis of the results of both searches would place
constraints on the numbers of Jovians that migrate, and thus on the
physical process itself.

Both microlensing and transit photometry have the ability to
push the discovery space for extrasolar planets into smaller mass
regimes than can be probed by the Doppler technique, over a
wider range of orbital radii.  Uranus- and Neptune-mass
planets orbiting several AU from their parent stars can
be detected with microlensing.  Less massive planets have caustics
so small that light curves probing them will be rare and suffer
from finite source size effects that dilute the signal,
unless smaller (and thus fainter) background stars
can be monitored (Bennett \& Rhie 1996; Gaudi \& Sackett 2000).
The best ground-based transit photometry ($\sim$0.001~mag) is sensitive
to planets with radii one-third that of Jupiter, i.e.,
planets similar to Uranus and Neptune.
Unless ground-based efforts considerably more substantial and
precise can be undertaken, the sensitivity to terrestrial-mass
planets promised by both techniques will await proposed space
missions such as COROT, Eddington, Kepler, MONS, MOST and GEST.
(For the latest status of these and other exoplanet projects,
see http://www.obspm.fr/planets.)

\acknowledgments

The PLANET team is grateful to the observatories that support their
research through generous allocations of telescope time, and to
Dr.~Lisa Germany for her observing assistance in generating
some of the first PLANET team transit data.

\end{document}